\documentclass[a4paper,superscriptaddress,twocolumn,preprintnumbers,showpacs,amsmath,amssymb,prl]{revtex4}
\usepackage{graphicx}
\usepackage{latexsym}
\usepackage{mathrsfs} 

\begin{document}
\title{Slow and remanent electric polarization of adsorbed BSA layer\\ evidenced by neutron reflection}

\author{A. Koutsioubas}
\affiliation{Laboratoire L\'eon Brillouin, CEA/CNRS UMR 12, CEA-Saclay, 91191 Gif-sur-Yvette cedex, France}

\author{D. Lairez}
\email[Corresponding author. E-mail: ]{lairez@cea.fr}
\affiliation{Laboratoire L\'eon Brillouin, CEA/CNRS UMR 12, CEA-Saclay, 91191 Gif-sur-Yvette cedex, France}

\author{G. Zalczer}
\affiliation{Service de Physique de l'Etat Condens\'{e}, CEA-Saclay, 91191 Gif-sur-Yvette cedex, France.}

\author{F. Cousin}
\affiliation{Laboratoire L\'eon Brillouin, CEA/CNRS UMR 12, CEA-Saclay, 91191 Gif-sur-Yvette cedex, France}

\begin{abstract}Using  neutron reflectivity together with an appropriate electrochemical cell, we have studied the effects of transverse electric field on the Bovine Serum Albumin (BSA) monolayer initially adsorbed at the interface of the aqueous solution and a conductive doped-silicon wafer. Depending on the sign of the initial potential, a second layer is adsorbed or not on top of the first whereas a subsequent reversal of potential has no effect. We show that this behaviour reveals the slow and remanent electric polarization of the first BSA layer. Based on the permanent dipolar structure of BSA, we suggest an analogy with dipolar glasses that may account for the slowness and memory of the process.
\end{abstract}

\date{29 Nov 2011} 

\maketitle

Protein adsorption on solid surfaces is a fundamental phenomenon that plays a central role in biotechnology; e.g. for biomaterials and biocompatibility improvement\cite{Nakanishi:2001kx}. Proteins are polyampholytes (i.e. carry both positive and negative charges) and amphiphiles (i.e. carry both hydrophilic and hydrophobic parts) and these features make the thermodynamics of their interactions with a surface particularly rich and extensively studied\cite{Norde:2008}. Studies of the dynamics of adsorbed proteins mainly focus on adsorption kinetics that may take place in a few tens of minutes\cite{Azzam:1977} but sometime is much slower, history dependent and displays some irreversible features\cite{mura:1991,Calonder:2001}. In these latter cases, the slow dynamics is viewed as inherent to the random addition of new molecules onto the surface and described in terms of the Random Sequential Adsorption (RSA) model\cite{Evans:1993bv} that displays slow dynamics typical of glass or jamming transition. The counterpart of this jammed behaviour is the slow relaxation of the in-plane density of the adsorbed layer after an external perturbation or spontaneous fluctuations\cite{Fadda:2009}. Much less studied and also presumably quite different are external perturbations that retain the density constant but should in principle operate on the orientation of adsorbed molecules, although this handling of adsorbed layer has many practical significances, e.g. for immobilized enzymes, immunoassays and biosensors. Due to the electrical charges of proteins, electrical stimuli are the most obvious with respect to this orientation. But studies in this direction are still sparse\cite{Morrissey1976,Kawaguchi:1988,Moulton2003,Beaglehole1998} and only concerned up to now with ellipsometry or electrochemical measurements, which produce ambiguous results due to technical limitations\cite{Kleijn:2004uq}.

In this paper, we report a neutron reflectivity study on the effects of transverse electric field on a Bovine Serum Albumin (BSA) monolayer initially adsorbed at silicon/water interface. Our results show a slow and remanent polarization of this layer revealed by the formation of a second layer on top of the first. Based on the BSA structure that displays a permanent electric dipole\cite{Seyrek:2003}, these features suggest an analogy with spin glasses that has not been considered until now.

\section{Samples and experimental device}
BSA (Sigma-Aldrich) was diluted at a concentration of 1\,mg/ml in heavy water (99.90 \%$^2$H, Euriso-top) with NaCl salt at a concentration $c_{\text{NaCl}}=0.02$\,M. The pD of solutions (potential of hydrogen $^2$H) was measured before and after the addition of protein and was found equal to 6.5, i.e. above the isoelectric point of BSA equal to 4.7.

Doped (n-type) optically flat silicon wafers of 1-10\,$\Omega$cm resistivity, were used as substrates for the  adsorption of proteins. Before each experiment, the silicon surface was cleaned by immersion in "piranha solution" (H$_2$SO$_4$ 70\%, H$_2$O$_2$ 30\%) for 10\,min and subsequent copious rinse with ultrapure water and a final rinse with heavy water.

\begin{figure}[!hbtp]
\centering
\includegraphics[width=1\linewidth]{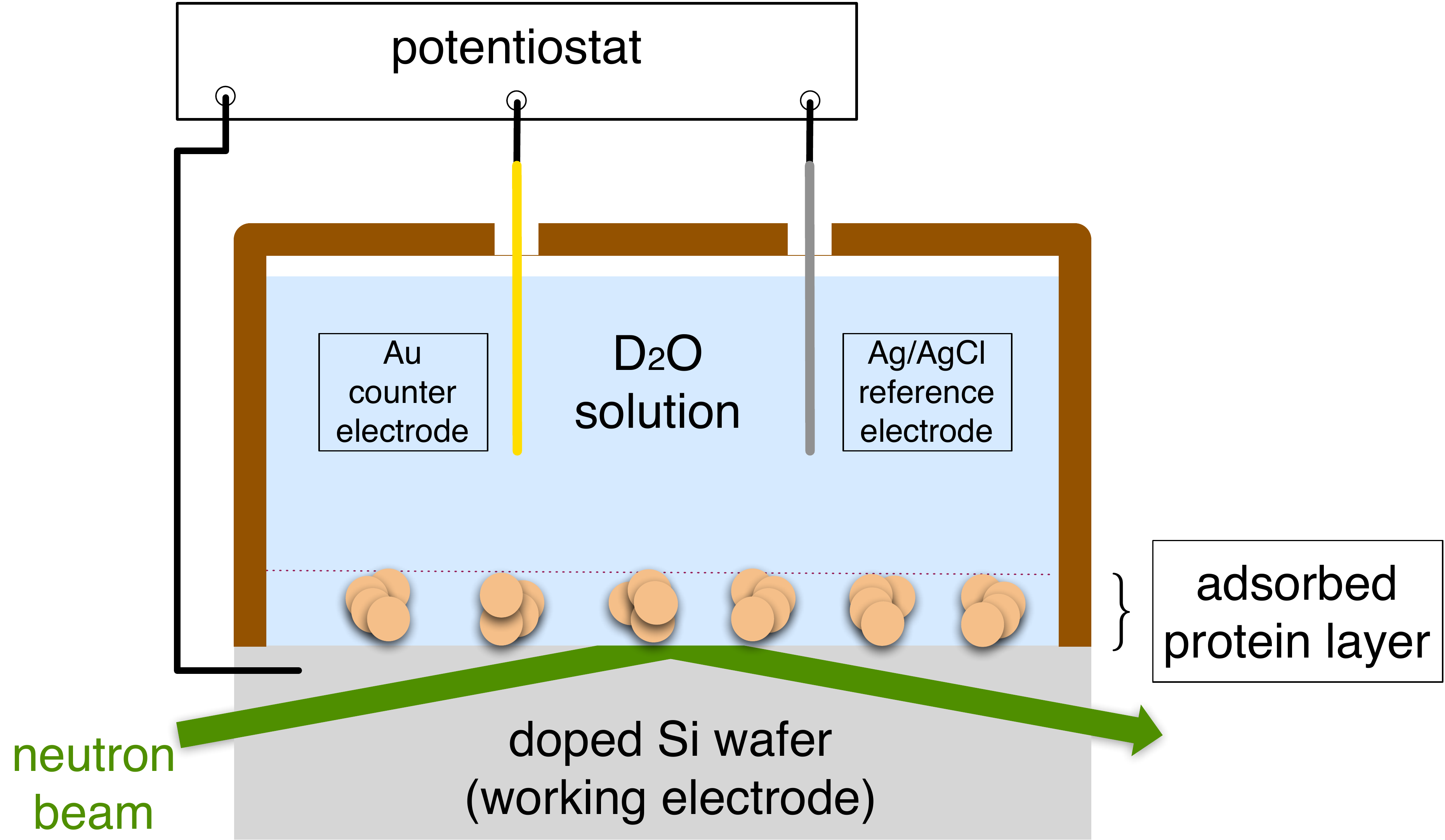}
\caption{Schematic layout of the device used for neutron reflectivity measurements.}
\label{figdevice}
\end{figure}

Neutron specular reflectivity measurements were performed at the EROS spectrometer at the Laboratoire L\'eon Brillouin. (incidence angle $\theta=1.34^\circ$, wavelength $3<\lambda<30$\,\AA). The neutron beam enters through the one side of the silicon wafer, is reflected at the silicon-solvent interface and exits from the other side of the wafer. A liquid cell is mounted on the top of the wafer for the immersion of its surface (Fig.\ref{figdevice}). The voltage is applied using a custom made potentiostat. The silicon wafer serves as the working electrode, a gold wire as the counter electrode, while a Ag/AgCl electrode is the reference. The applied potential range of $\pm 1.0$\,V is inside the electrochemical window of water equal to $\pm 1.23$\,V. Thus this chosen voltage ensures maximum effect without water electrolysis. Actually, gas formation near the electrode was not observed. In addition, the analysis of neutron reflectivity spectra for bare silicon wafers in contact with water under applied potentials showed no visible change over a period of 24 hours.

The stability of the bulk BSA solution with time and upon applied voltage was also checked by circular dichroism measurements in the bandwidth 200-250\,nm performed with a JASCO J-815 CD spectrometer (CEA/DSV/iBiTec/SIMOPRO).

\section{Results}

\subsection{Measurement procedure and raw data examination}
The acquisition of a reflectivity spectrum takes at least one hour, which is already long in our system compared to the initial adsorption kinetics that takes place in a few tens of minutes. Actually, the first  spectrum measured on a wafer in contact with the protein solution clearly shows an additional layer compared to the bare wafer. However, no further evolution monitored for several hours has been observed prior to the voltage was applied. Then, this initial adsorbed layer was monitored for several hours under applied voltage. Two different voltage sequences were considered: 
\begin{enumerate}
\item 0/+1/-1\,V: that corresponds to 0\,V for 24 hours then +1\,V for 24 hours then -1\,V for 24 hours;
\item 0/-1/+1\,V.
\end{enumerate}
These voltage values were chosen in order to expect the maximum effect without water hydrolysis. For each voltage sequence, the measurement cell was kept in place on the spectrometer without any change except for the voltage variations.

\begin{figure}[hbtp]
\centering
\includegraphics[width=1\linewidth]{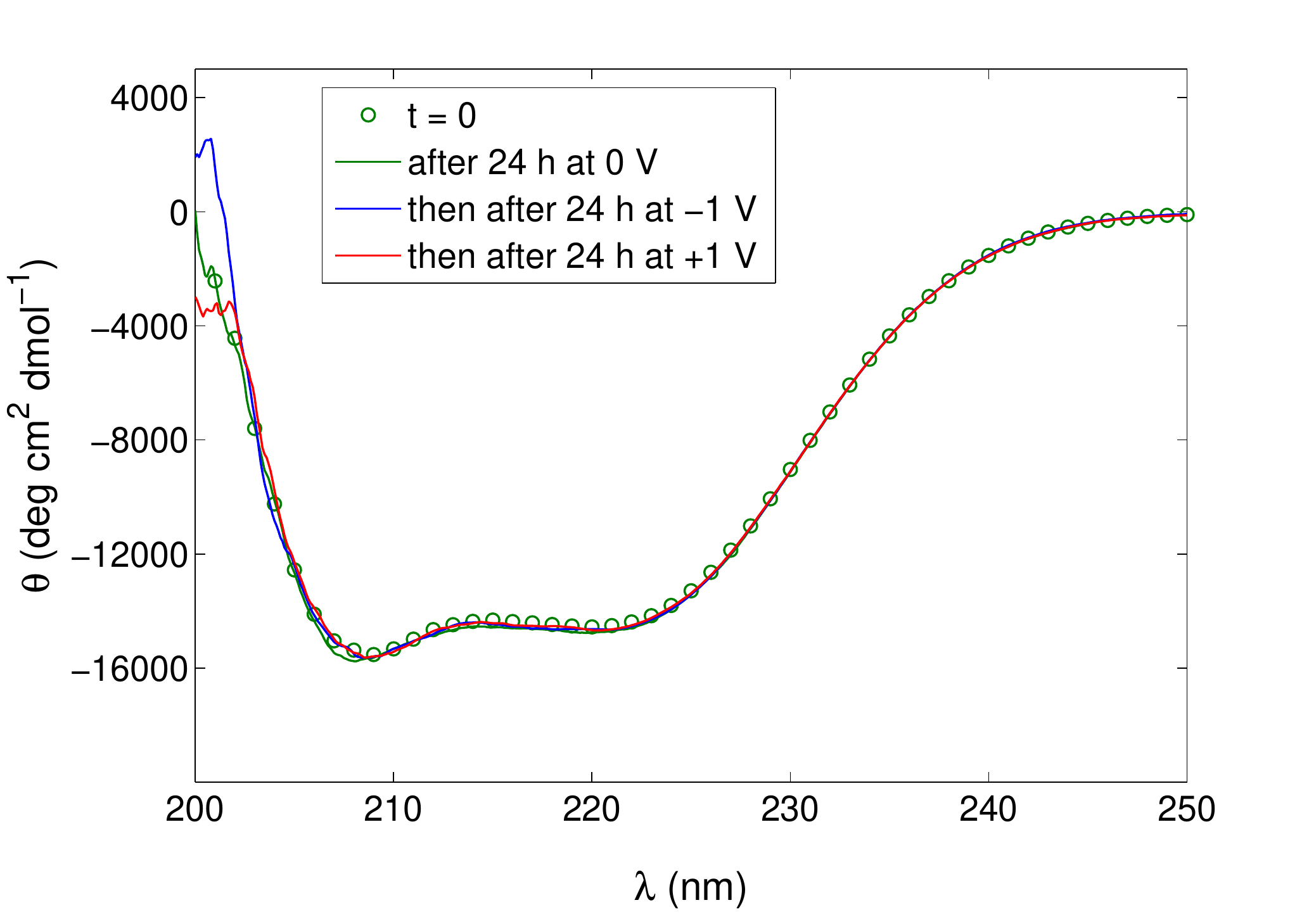}
\caption{Circular dichroism ellipticity $\theta$ per amino acid concentration unit of the BSA bulk solution for the voltage sequence number 2 as a function of the wavelength $\lambda$.}
\label{figcd}
\end{figure}

As we are concerned with long time kinetics, the stability of the bulk BSA solution is questionable.
To address this point, we performed circular dichroism (CD) measurements on samples taken in the bulk solution. Actually, the secondary structure of BSA contains 50\% of random conformation and 50\% of $\alpha$-helix. In addition, the ternary structure of BSA is highly stabilized by up to 17 disulfide bonds. In this context, CD spectra in the bandwidth 200-250\,nm (that signs the total $\alpha$-helix amount) is suitable to check the protein integrity\cite{Norde:2000fk}. Results are plotted in Fig.\ref{figcd} for the voltage sequence number 2. No significant change of CD spectra are observed.

\begin{figure}[hbtp]
\centering
\includegraphics[width=1\linewidth]{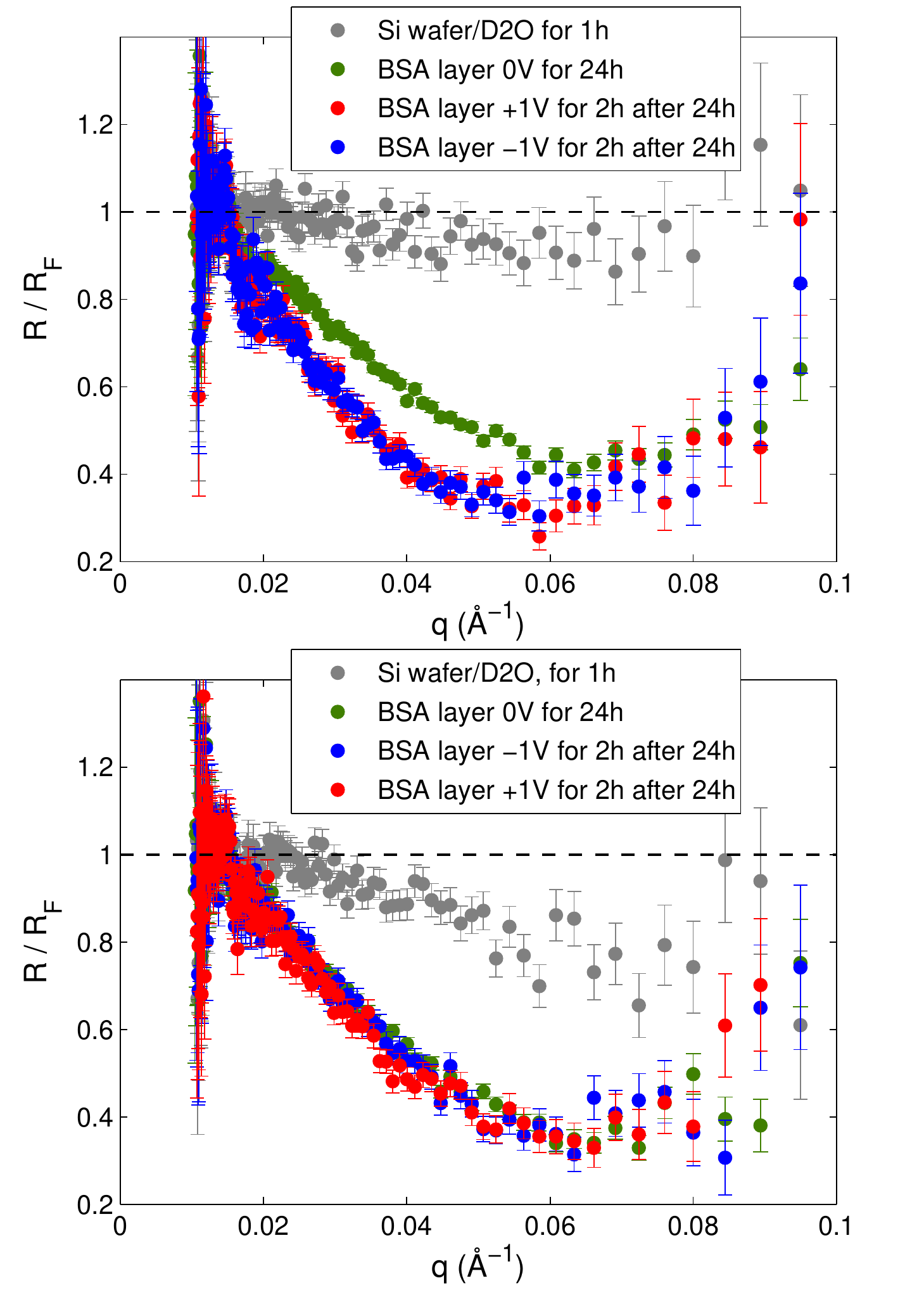}
\caption{Ratio $R/R_F$ vs scattering vector $q$, where $R$ is the reflectivity of adsorbed BSA layer or bare wafer and $R_F$ the calculated Fresnel reflectivity of an ideal silicon-solvent interface. The two figures correspond to two different sample histories i.e. two voltage sequences. Top: 0/+1/-1\,V sequence;  Bottom: 0/-1/+1\,V sequence. For 0\,V (green symbols), the data acquisition time was equal to 24\,h as no evolution kinetics of the spectrum was observed. For +1 and -1\,V, (red and blue symbols), spectra were acquired for 2\,h after a waiting time of 24\,h at this voltage.}
\label{rawcurves}
\end{figure}

In Fig.\ref{rawcurves}, the reflectivity spectra $R$ as a function of the scattering vector modulus $q=4\pi \sin(\theta) / \lambda$, obtained for these two independent voltage sequences are compared to the reflectivity of the bare wafer. In this figure, the spectra are divided by the Fresnel reflectivity $
R_F(q)=\left|{\left({q-(q^2-q_c^2)^{1/2}}\right)/\left({q+(q^2-q_c^2)^{1/2}}\right)}\right|^2
$
of the silicon-solvent interface, where $q_c$ is the critical value of $q$ below which total reflectivity occurs. In this representation, deviations from unity (dashed line in Fig.\ref{rawcurves}) indicate additional contributions to this ideal interface. For the bare wafer (grey points in Fig.\ref{rawcurves}), this deviation is due to the native SiO$_2$ layer that may differ depending on the wafer (this explains the different spectra for the two bare wafers in Fig.\ref{rawcurves}). For the wafer in contact with the BSA solution, the ratio $R/R_{F}$ deviates even more from unity (colored points in Fig.\ref{rawcurves}) indicating BSA adsorption that occurs even without any voltage (green points). Once the voltage is applied, the evolution of the reflectivity differs depending on the voltage sequence. When the positive voltage is imposed first there is an appreciable change of the reflectivity, that implies an overall increase of the adsorbed amount. On the contrary, when the negative voltage is imposed first, the adsorbed BSA layer seems unaffected. This phenomenon is the core of this paper. Without further data analysis or fitting, it clearly indicates that the adsorbed BSA layer:
\begin{itemize}
\item is sensitive to the electrical field;
\item is sensitive to the polarity of the field;
\item has a memory and history dependent behaviour (as $0\rightarrow+1$\,V does not yield the same final result as $0\rightarrow-1\rightarrow+1$\,V).
\end{itemize}
These points are detailed and discussed in the following.

\subsection{Data fitting procedure}

\begin{figure*}[!htbp]
\centering
\begin{tabular}{lll}
\includegraphics[width=0.32\linewidth]{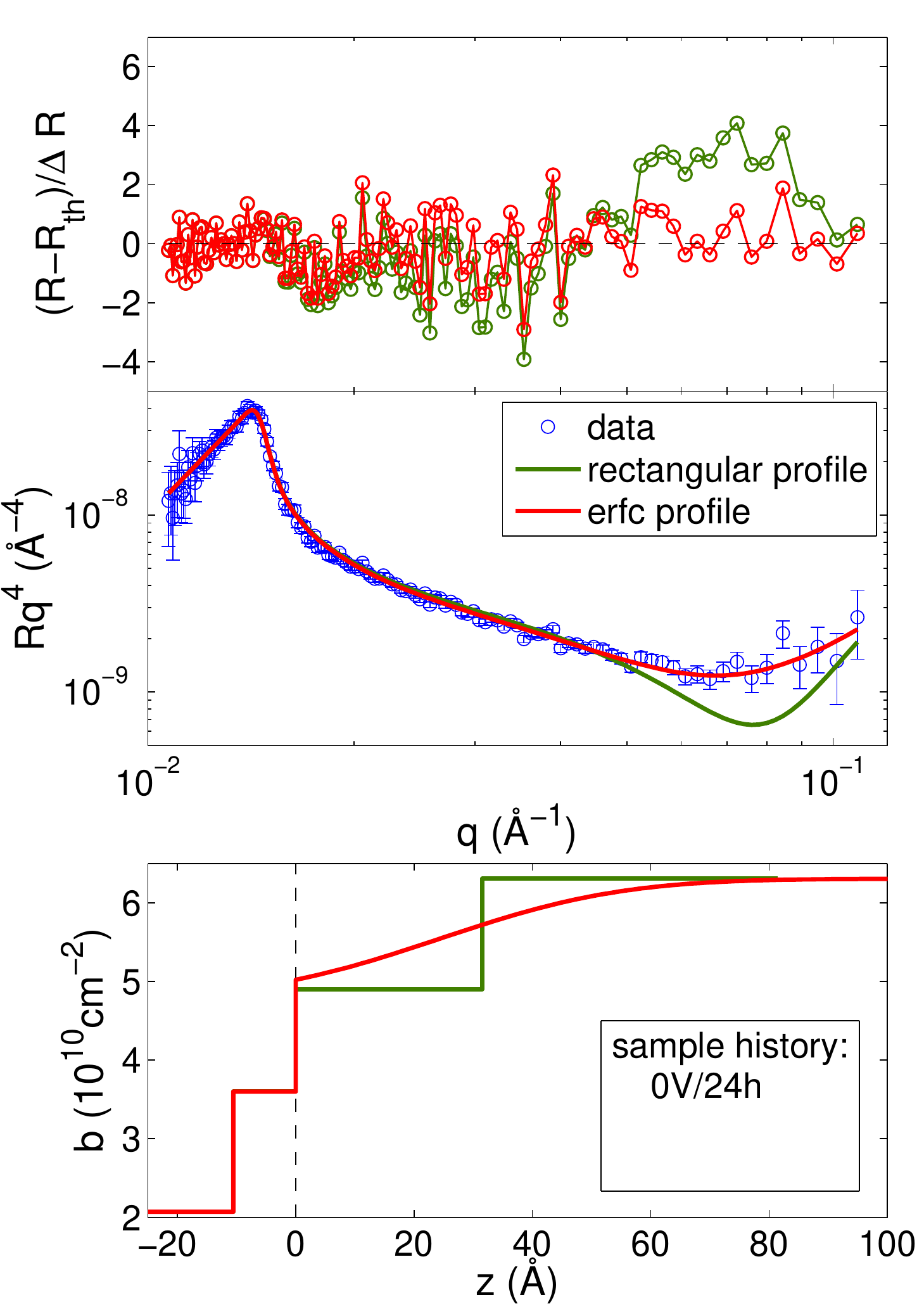}&
\includegraphics[width=0.32\linewidth]{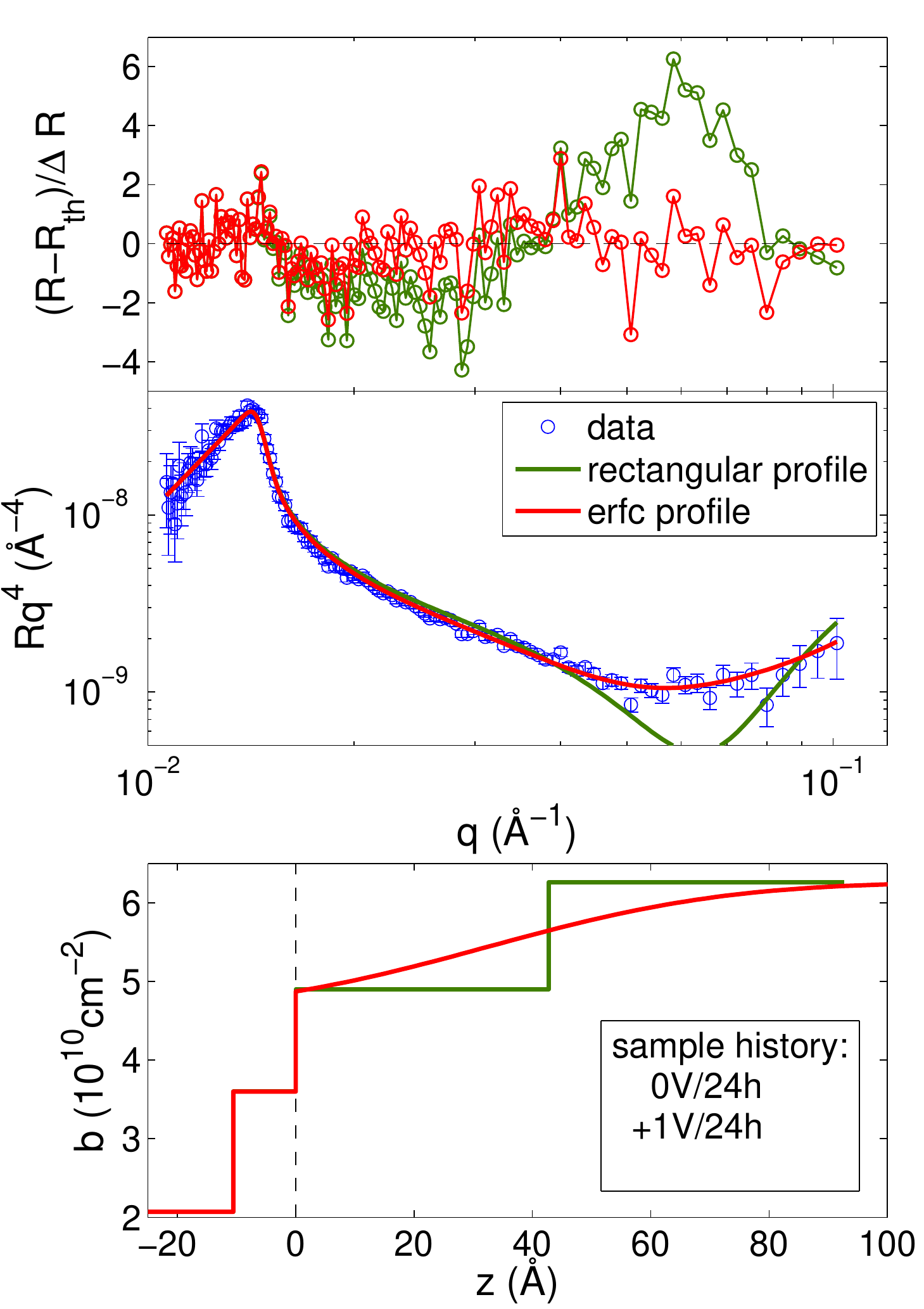}&
\includegraphics[width=0.32\linewidth]{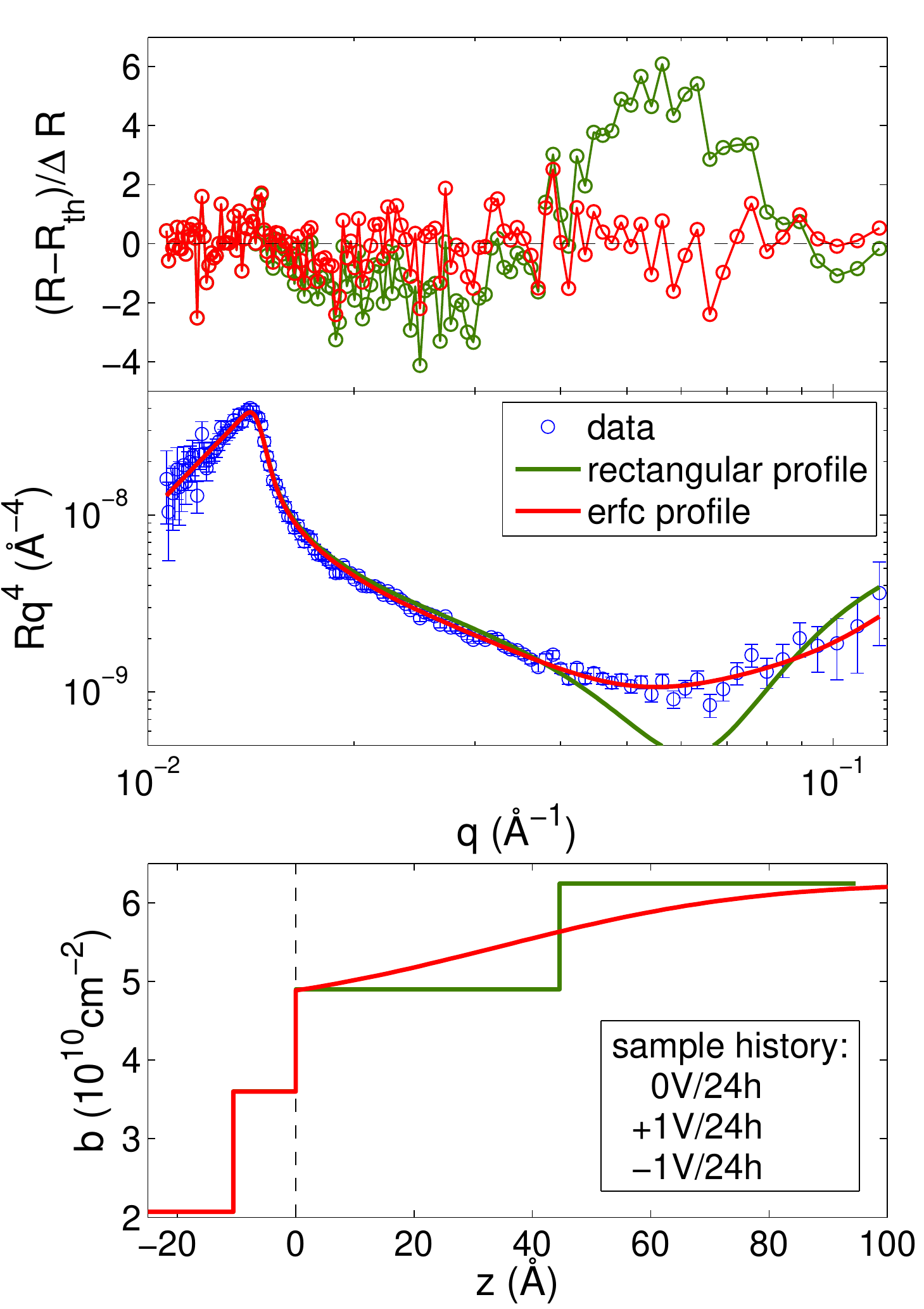}\\
\textbf{a)}&\textbf{b)}&\textbf{c)}
\end{tabular}
\caption{Reflectivity of adsorbed BSA layer (acquired during 2\,h): \textbf{a)} measured after 24\,h at 0V; \textbf{b)} then  after 24\,h at +1\,V; \textbf{c)} then after 24\,h at -1\,V. Middle: spectra and best fits using erfc and rectangular profiles. Top: fits residuals. Bottom: scattering length density profile corresponding to the best fits ($z=0$ is the wafer surface position). Wafer characteristics determined from the independent measurement of the wafer in contact with D$_2$O: $\tilde b_{\textrm{Si}}=2.07\times 10^{10}$\,cm$^{-2}$, oxide layer: $\tilde b_{\textrm{Si0$_2$}}=3.6\times 10^{10}$\,cm$^{-2}$, thickness  $10.5\pm0.7$\,\AA. Scattering length density of the solvent determined from the $q_c$ value of the reflectivity: \textbf{a)}~$\tilde b_s=6.31\pm0.01\times 10^{10}$\,cm$^{-2}$; \textbf{b)}~$6.28\pm0.01\times 10^{10}$\,cm$^{-2}$; \textbf{c)}~$6.26\pm0.01\times 10^{10}$\,cm$^{-2}$. The whole spectra were fitted with the scattering length density $\tilde b_p(z=0)$ and the thickness $h$ of the protein layer as  the only two free parameters that are found as equal to: 
\textbf{a)}~$\tilde b_p(0)=(5.02\pm0.05)\times 10^{10}$\,cm$^{-2}$, $h=(24.3\pm0.7)$\,\AA, $\chi^2=0.85$ 
(erfc), $\tilde b_p(0)=(4.9\pm0.2)\times 10^{10}$\,cm$^{-2}$, $h=(31\pm2)$\,\AA, $\chi^2=2.1$ (rectangular);
\textbf{b)}~$\tilde b_p(0)=(4.87\pm0.03)\times 10^{10}$\,cm$^{-2}$, $h=(32.3\pm0.6)$\,\AA, $\chi^2=0.98$ (erfc),
$\tilde b_p(0)=(4.9\pm0.1)\times 10^{10}$\,cm$^{-2}$, $h=(42.7\pm1.5)$\,\AA, $\chi^2=3.65$ (rectangular).
\textbf{c)}~$\tilde b_p(0)=(4.89\pm0.02)\times 10^{10}$\,cm$^{-2}$, $h=(34.1\pm0.5)$\,\AA, $\chi^2=0.82$ 
(erfc), $\tilde b_p(0)=(4.9\pm0.1)\times 10^{10}$\,cm$^{-2}$, $h=(45\pm1.5)$\,\AA, $\chi^2=3.94$ 
(rectangular).}
\label{fits1}
\end{figure*}

The specular reflectance $R(q)$ is related to the average density profile of the scattering length $\tilde b(z)$ normal to the wafer surface. $\tilde b(z)$ was determined through least squares fitting of $R(q)$ performed with the following method. Theoretical reflectance $R_{\textrm{th}}$ was convoluted by the $q$-dependent spectrometer resolution and optimized in order to minimize the relative distance per point between $R_{\textrm{th}}$ and $R$ defined as: $\chi^2=\sum_{i=1}^N\left({(R_i-R_{\textrm{th i}})}/\Delta R_i\right)^2/N$, where $N$ is the number of data points and $\Delta R$ the statistical error for each point deduced from the neutron counting. All spectra were fitted fixing the scattering length density of the silicon block to $\tilde b_{\textrm{Si}}=2.07\times 10^{10}$\,cm$^{-2}$. For each experiment, the scattering length density $\tilde b_s$ of the solvent was determined by fitting the spectrum by the Fresnel reflectivity in the region of the critical value $q_c=(16\pi (\tilde b_s-\tilde b_{\textrm{Si}}))^{1/2}$. The optimal value for $\tilde b_s$ was found to lie between $6.2\times 10^{10}$ and $6.4\times 10^{10}$\,cm$^{-2}$ depending on the experiment (D$_2$O contains a small amount of H$_2$O that may vary with the experiment and increase with time) and was kept constant in the rest of the fitting procedure. Theoretical reflectances of multilayers were calculated by the transfer-matrix method~\cite{Abeles:1950}. For each experiments series, the thickness of the native oxide layer was determined by fitting the spectrum of the bare wafer in contact with the solvent, fixing the scattering length density of the oxide ($\tilde b_{\textrm{Si0$_2$}}=3.6\times 10^{10}$\,cm$^{-2}$) and assuming no roughness of the interfaces. The so determined characteristics of bare wafers were kept constant to analyze the additional contribution $\tilde b_p(z)$ of the adsorbed protein layer to the overall density profile. A first attempt to fit these spectra was done using a simple rectangular profile for $\tilde b_p(z)$, but results obliged us to consider a softer profile in order to account for the roughness of the protein-layer/solvent interface. Traditionally, this is achieved by multiplying the reflectivity of the interface by a Debye-Waller factor of the form $e^{-q^2\sigma_h^2/2}$, where $\sigma_h$ is the standard deviation of the interface height $h$. This method to fit our data yields a protein-layer thickness $h$ and roughness $\sigma_h$ of the same order of magnitude. To overcome this unphysical result we used the following profile:
\begin{equation}\label{profil}
\begin{array}{l}
\tilde b_p(z\le0)=0\\
\tilde b_p(z>0)=(\tilde b_p(0)-\tilde b_s)\frac{\text{erfc}\left({ \frac{z-h}{\sqrt{2}h}}\right)}{\text{erfc}\left({-1/\sqrt 2}\right)}+\tilde b_s
\end{array}
\end{equation}
that is mathematically equivalent to a Debye-Waller factor with $\sigma_h=h$, except for the cutoff for $z\le0$. The fitting procedure was the following: $\tilde b_p(z)$ was discretized into 100 layers extending to $b_p(z)/b_p(0)=10^{-3}$; these layers were added to those of the bare wafer and $R_{\textrm{th}}$ calculated by the transfer-matrix method; the two parameters $h$ and $\tilde b_p(0)$ were optimized to minimize $\chi^2$. In Fig.\ref{fits1}, for the voltage sequence 0/+1/-1\,V, three typical spectra are plotted in the representation $Rq^4$ as a function of $q$ with the best fits obtained for rectangular and erfc profiles.

\subsection{Kinetics}

\begin{figure}[!htbp]
\centering
\includegraphics[width=1\linewidth]{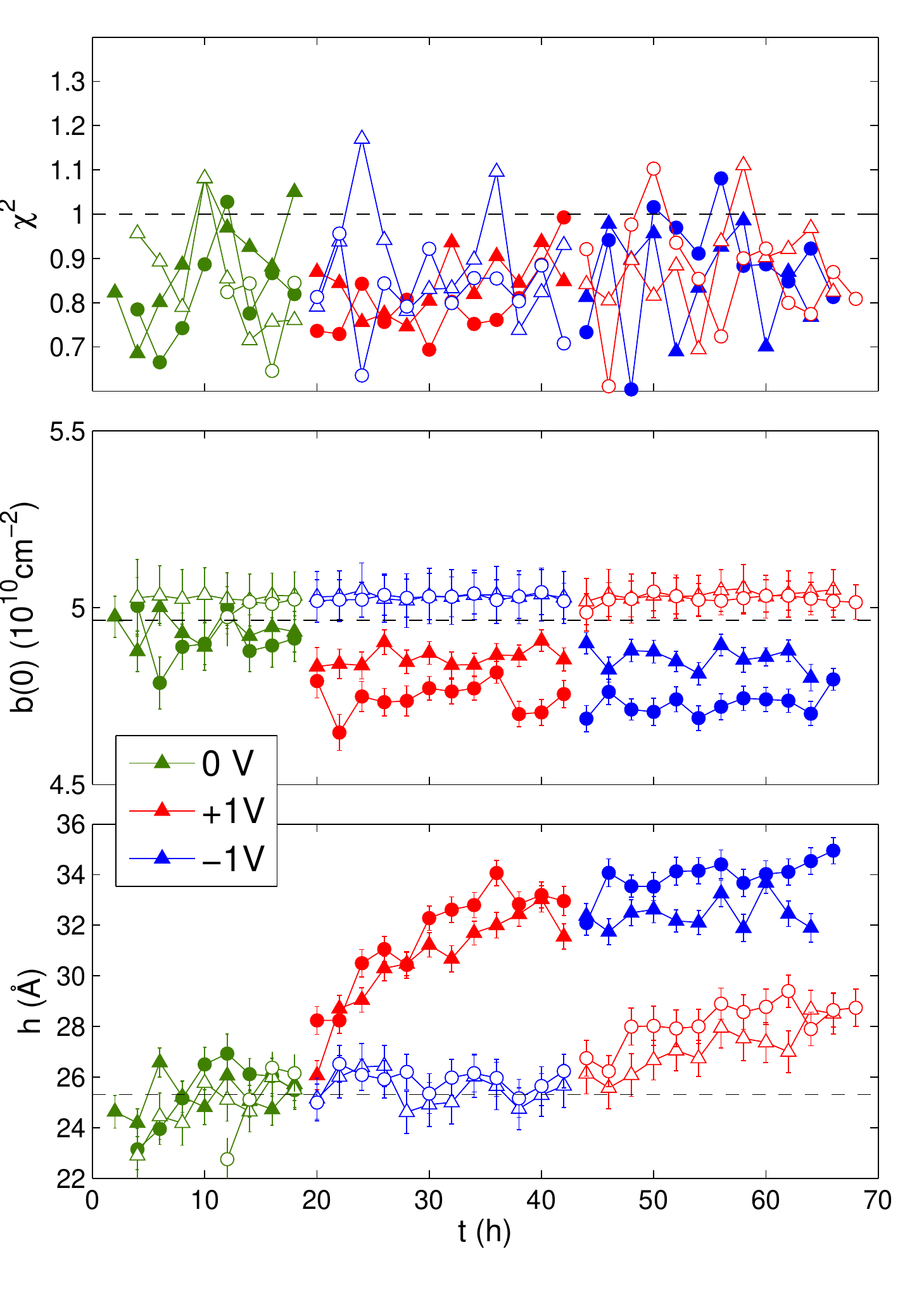}
\caption{Variation with time $t$ of the thickness $h$ (bottom) and scattering length density $\tilde b(0)$ (middle) of the BSA layer for the two voltage sequences deduced from best fits of reflectivity spectra with a density profile following Eq.\ref{profil}. Full symbols: 0/+1/-1\,V sequence; open symbols: 0/-1/+1\,V sequence.
Dashed lines correspond to mean values at 0\,V (bottom and middle) and $\chi^2=1$ (top).}
\label{results}
\end{figure}

The two parameters erfc profile of Eq.\ref{profil} (that gives the best results for the fitting procedure) has been retained to fit the spectra recorded all along the voltage and history dependent adsorption kinetics. Each spectrum was acquired for 2\,h. For all spectra, the fitting procedure leads to $\chi^2$ values lying between 0.7 and 1.
Results for the scattering length density $\tilde b_p(0)$ and thickness $h$ of the protein layer are plotted in Fig.\ref{results} as a function of elapsed time for the two voltage sequences 0/+1/-1\,V, and 0/-1/+1\,V. Each voltage sequence has been repeated with a clean buffer and a new protein solution yielding to a good reproducibility of the observed phenomenon.

\section{Discussion}

\subsection{Protein volume fraction at the interface}
Whatever the voltage sequence (Fig.\ref{results}), the scattering length density $\tilde b_p(0)$ of the protein layer at the surface of the wafer is practically constant and independent of the voltage for a given experiment and also nearly independent of the experiment. At 0\,V the mean value is
$<{\tilde b_p(0)}>_{\text{0\,V}}=(4.97\pm 0.06)\times 10^{10}\text{\,cm}^{-2}$.
$\tilde b_p$ is related to the volume fraction $\phi$ occupied by the protein in the adsorbed layer through the relation: 
$\phi(z)=(\tilde b_p(z)-\tilde b_s)/(\tilde b_{\text{BSA}}-\tilde b_s)$,
where $\tilde b_{\text{BSA}}=2.97\times 10^{10}$\,cm$^{-2}$ is the scattering length density of BSA calculated by assuming that  only $80\%$ of all labile protons of the protein are exchanged~\cite{Jacrot:1976} (this classical value comes from some labile protons that remains buried inside protein globule without any contact with the solvent). The mean value $<{\tilde b_p(0)}>_{\text{0\,V}}$ gives
$<{\phi(0)}>_{\text{0\,V}}=0.40\pm 0.02$. The small decrease of $\tilde b_p(0)$ observed for the $0/+1/-1$\,V sequence corresponds to ${\phi(0)}=0.47\pm 0.02$. We believe this difference is insignificant and probably due to the shape of the density profile chosen for data fitting, which becomes inappropriate. Note that this non compact surface covering of proteins is consistent with the shape of the density profile that involves a surface roughness of the layer of the order of magnitude of its thickness. In addition, these volume fraction values are very close to the maximum packing fraction of randomly adsorbed spheres that is found of the order of 0.5 within the 2D-RSA model~\cite{Williams_2003}. 

\subsection{Thickness of the protein layer}
Whereas $\phi(0)$ remains roughly constant, the voltage sequence affects the thickness of the protein layer: at 0\,V, $<h>_{\text{0\,V}}=(25\pm1)\text{\,\AA}$, this value slowly increases up to 35\,\AA\ by applying +1\,V first, but remains nearly constant if -1\,V is applied first. In order to be compared to previous works, the surface excess $\Gamma$ is suitable because its estimation is model independent: $\Gamma=\rho_p\int_0^\infty \phi(z) dz$, where $\rho_p=1.35$\,g/cm$^3$ is the density of globular proteins~\cite{Creighton:1997}. Note that as the variation of ${\phi(0)}$ is small in our case, the variation of the surface excess follows the variation of the characteristic thickness $h$. The mean value measured at 0\,V is $<{\Gamma}>_{\text{0\,V}}=(1.6\pm 0.1)$\,mg/m$^2$. By applying +1\,V, this value increases up to 2.4\,mg/m$^2$. This range of values is fully compatible with previous neutron reflectivity studies of BSA adsorption\cite{Su:1998,Lu:2007uq}.

The crystallographic structure of BSA is still unknown. However, BSA has the same molecular weight as Human Serum Albumin (HSA) and 76\% sequence identity\cite{Peters-Jr.:1985fk}. The structure of the latter, available on the Protein Data Bank website (http://www.rcsb.org/pdb/) under the identifier 1AO6, allows us to calculate its radius of gyration as equal to $R_g=27\text{\,\AA}$, in very good agreement with measurements on BSA solutions by small angle scattering~\cite{Bendedouch:1983, Zhang:2007}. In most cases once adsorbed on solid, globular proteins are partly denatured, however they still keep a compact and globular conformation\cite{Norde:2008}. This is experimentally confirmed for BSA at silica-water interface\cite{Lensun:2002uq,Larsericsdotter:2005fk}. It can be reasonably assumed that this partly denatured globule retains the same radius of gyration. The mean thickness of the initial BSA layer measured at 0\,V is thus comparable with $R_g$ rather than with the protein diameter. In our case, this is due to the soft density profile (Eq.\ref{profil}) that defines a characteristic thickness of the layer obviously smaller than its full extent. Thus, the value here reported for the thickness of the initial layer adsorbed at the wafer surface prior to voltage application is fully consistent with a BSA monolayer.

\subsection{Electric field}

In our experimental device and at our working voltage, working and counter electrodes are "irreversible electrodes" in the electrochemical sense, i.e. there is no red-ox reaction at the interface and no ion-electron transfer. Thus in our case, due to this high resistance of the interfaces of the electrodes most of the potential drop occurs across the ions double layer (Stern-Gouy-Chapman) at each electrode surface, so that the electric field within the conducting electrolyte solution is negligible. 

\begin{figure}[!htbp]
\centering
\includegraphics[width=1\linewidth]{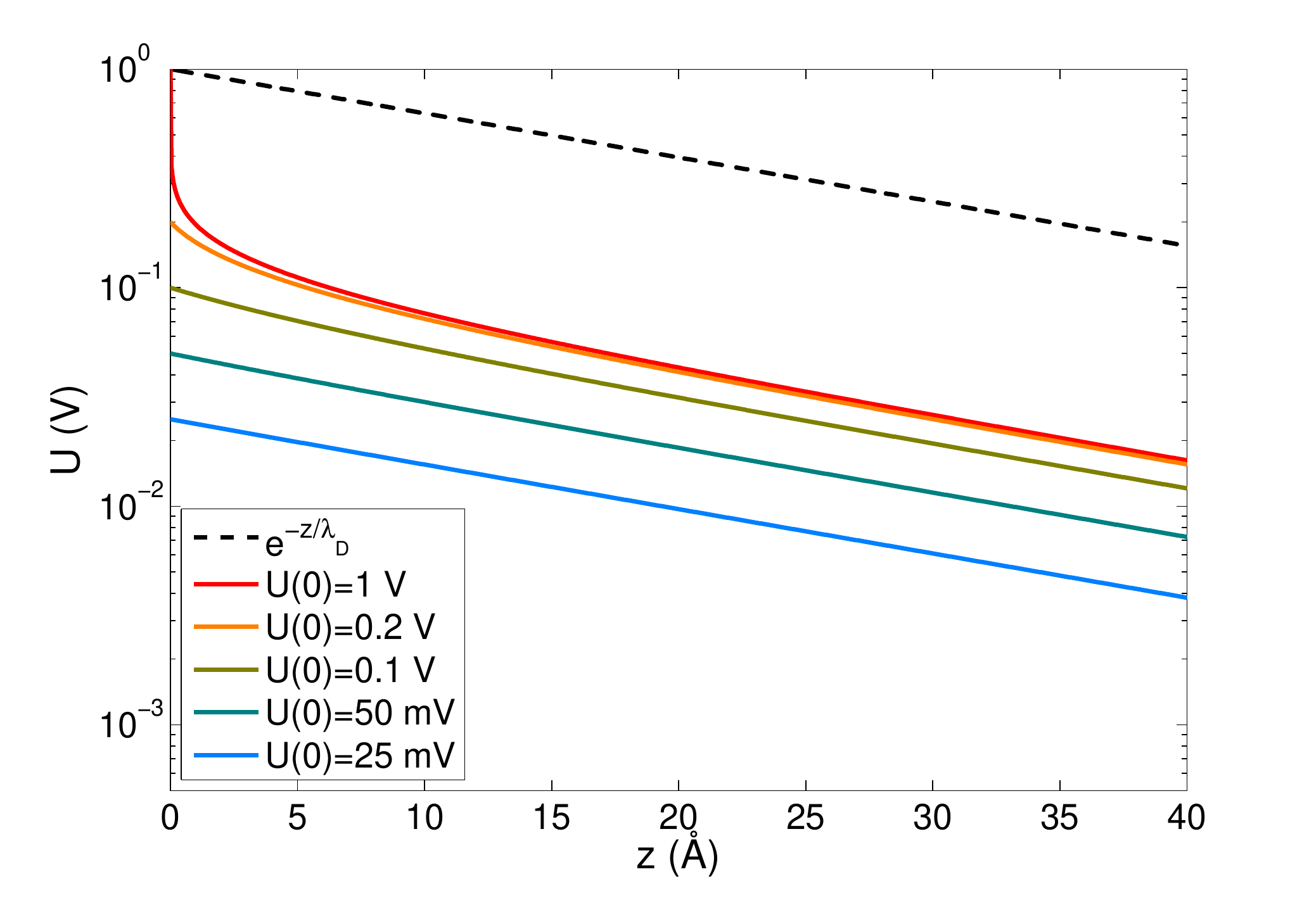}
\caption{Solid lines: electric potential $U(z)$ calculated following Poisson-Boltzmann equation\cite{Butt:2003} with $c_{\text{NaCl}}=0.02$\,M corresponding to $\lambda_D=21.5\text{\,\AA}$ and for different values of the surface potential $U(0)>25$\,mV, i.e. $U(0)>kT/\text{e}$. Dashed line: exponential decay $U(z)=e^{-z/\lambda_D}$.}
\label{pb}
\end{figure}

At neutral pH, BSA has a net charge number of $-18$\cite{Peters-Jr.:1985fk} and an effective charge number of $-10$ accounting for counter-ions condensation\cite{Bhme:2007uq}. However, whatever the sign of the applied voltage no desorption of the initial BSA layer is observed. Near the wafer surface, let us consider the diffuse layer of counter-ions (Gouy-Chapman layer) that extends over the Debye screening length of electrostatic interactions, $\lambda_D=(\epsilon kT/2\text{e}^2c_{\text{NaCl}})^{1/2}$, where $\epsilon$ is the dielectric permittivity, $kT$ the thermal energy, $\text{e}$ the electronic charge and $c_{\text{NaCl}}$ the salt concentration. Here $c_{\text{NaCl}}=0.02$\,M leads to $\lambda_D=21.5\text{\,\AA}$. 
At low surface potential $U(0)$ (i.e. $U(0)<kT/\text{e}\simeq 25\text{\,mV}$, the Poisson-Bolzmann equation can be linearized leading to $U(z)=U(0)e^{-z/\lambda_D}$.
However, at higher surface potential ($U(0)> 25\text{\,mV}$), exact calculations are needed\cite{Butt:2003}. Results for $U(z)$ are plotted in Fig.\ref{pb} for different surface potentials $U(0)$. 
One can see that for increasing surface potential, at the near-vicinity of the surface ($z\lesssim5\text{\,\AA}$), $U(z)$ separates from a simple exponential decay due to counter-ions condensation on the surface (Stern layer), but more interesting to understand our experiments is the saturation effect at longer distance from the surface. Actually, increasing the surface potential causes a densification of the Stern layer that in turn lets the potential unchanged at longer distances. In Fig.\ref{pb}, it appears that  for $U(0) \gtrsim 0.2$\,V and beyond 5\,\AA, $U(z)$ is independent of $U(0)$ and decays exponentially with an amplitude of the order of $u_0\simeq 0.1$\,V. For the sake of simplicity, let us write $U(z)=u_0e^{-z/\lambda_D}$. Assuming a uniform electrical charge density within the BSA layer of the form $q(z)=n\text{e}/2h$, with $n$ the effective charge number of BSA, one gets the  electrostatic potential energy: $W=\int_0^{2h}U(z)q(z)dz\simeq 0.4\text{\,eV}\simeq15$\,$kT$. As no desorption has been observed even for $U(0)=-1$\,V, the adsorption energy is thus higher than 15\,$kT$. This is in agreement with the adsorption free energy of the order of 20\,$kT$ reported for BSA at the water-silica interface\cite{oss:2001}. Note that the main contribution for this high adsorption energy is usually imputed to a gain of entropy (protein conformational entropy due to partial loss of structural elements and water entropy)\cite{Norde:2008,Besseling:1997fk}. 

\subsection{Effect of the polarity sequence}
Although no desorption has been observed, the BSA layer is clearly sensitive to the electrical field in the both sequences of voltages that have been studied:
\begin{enumerate}
\item When +1\,V is first imposed, the BSA layer presents a slow thickness augmentation with a characteristic time of the order of 10 hours, whereas the protein volume fraction $\phi(0)$ at $z=0$ remains constant,
\item When -1\,V is first imposed, the BSA layer remains apparently unaffected, but is clearly not in its initial state, because a subsequent reversal of voltage (-1$\rightarrow$+1\,V) has only a small (or even more slower) effect on adsorption compared to the sequence  0$\rightarrow$+1\,V. This is a clear memory effect.
\end{enumerate}

\begin{figure}[!htbp]
\centering
\includegraphics[width=0.8\linewidth]{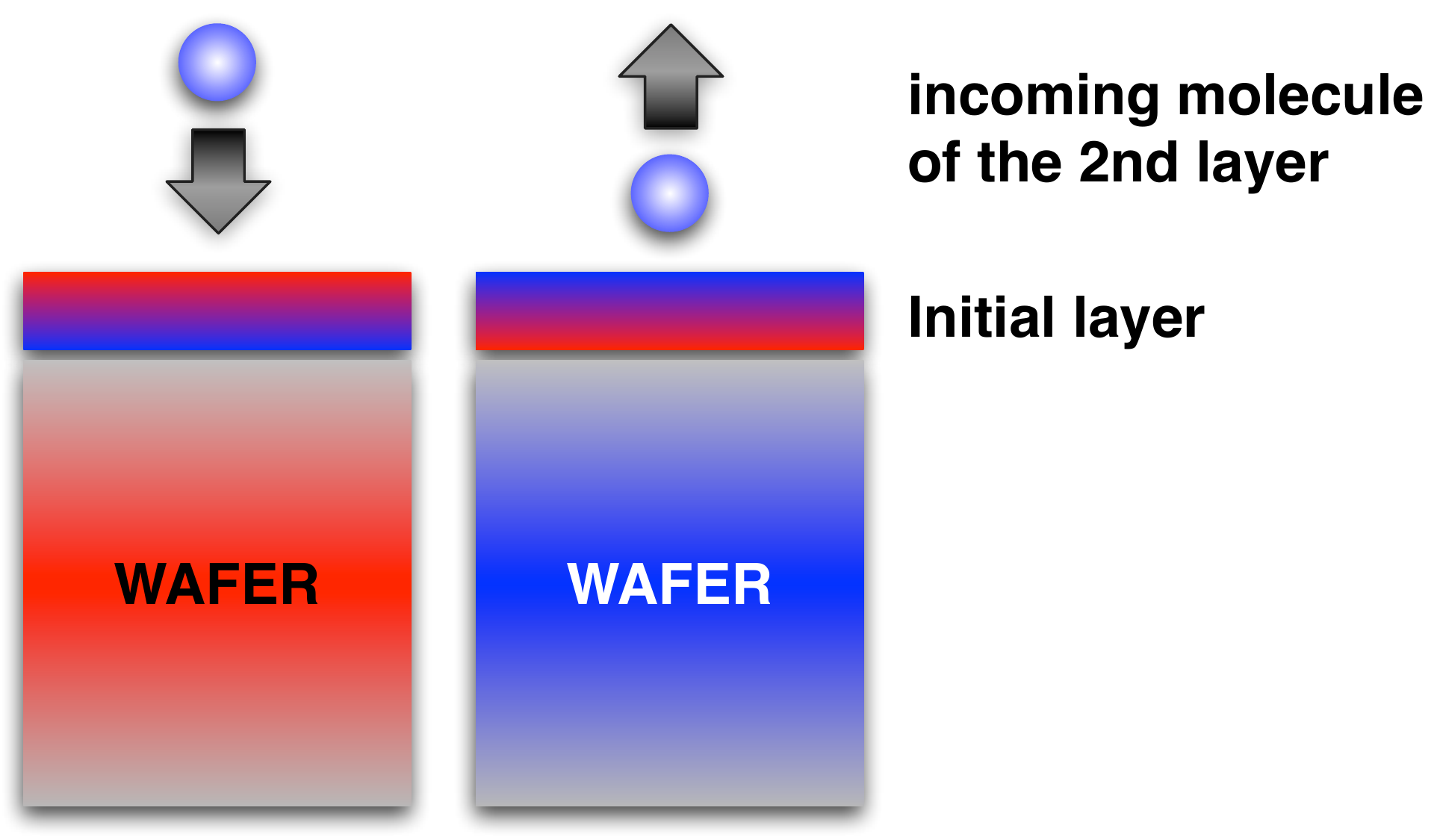}
\caption{Schematic view of the first layer polarization revealed by the formation of a second layer on top of the first. Red and blue correspond to positive and negative potential, respectively.}
\label{scenario}
\end{figure}

In the first case, our results can only be attributed to the formation of a second BSA layer on top of the first. 
However, the Debye screening length, which is in the bulk of the same order as the thickness of the first BSA layer, is even more reduced at the vicinity of the surface by the electric charges carried by the adsorbed proteins. So, the electric potential of the wafer surface is fully screened beyond the thickness of the first BSA layer. In other words, a random distribution of the electrical charges inside the first protein layer would result in proteins of the supernatant solution insensitive to the wafer voltage. Clearly, this is not the case. We argue that this result proves the field-induced electric polarization of the first BSA layer. Then, the resulting sign of the surface-charge of this first layer favors or prevents incoming free BSA molecules that carry a negative net charge. In this way a second layer is progressively formed or not (see Fig.\ref{scenario}). 

\subsection{Slowness and memory of the process}

Let us focus on the slowness of the process in sequence~1. It cannot be inherent to the random adsorption mechanism itself because: 1)~the adsorption of the first layer is already very rapid (i.e. less than the acquisition time of the first spectrum) and leads to a protein volume fraction near the jamming limit; 2)~the 50\% increase in the surface excess  when +1\,V is applied corresponds to a second layer volume fraction of the order of 0.2, i.e. far from the jamming limit, which implies an adsorption kinetics that should be even more rapid. Then, two points remain puzzling and have to retain our attention: 
\begin{enumerate}
\item the slowness of the polarization process (sequence~1). 
\item the memory effect (sequence~2).
\end{enumerate}

Actually at neutral pH, due to the spatial repartition of its electrostatic charges, BSA displays a permanent electric dipole\cite{Seyrek:2003}. Because this spatial arrangement is stabilized by up to 17 disulfide bonds\cite{Katchalski:1957fk}, it can be reasonably assumed that BSA molecules still keep this overall dipolar structure when adsorbed on silica, or under an electrical field, even if they are partly denatured at small length scale\cite{Lensun:2002uq,Larsericsdotter:2005fk,Zhu:2000}. Thus at a molecular level, the electric polarization of the first BSA layer could be due to the orientation of these permanent dipoles. One may expect that BSA rotation has practically no effect on its conformational entropy nor on water entropy (that have a major contribution to the adsorption energy), so that BSA should be free to rotate without desorbing. However, due to the high volume fraction of BSA in this first layer, dipole-dipole interactions cannot be neglected and the system should come under a 2D classical Heisenberg model with a transverse field and first neighbors antiferro-coupling between randomly positioned sites resulting from random sequential adsorption (RSA) of the first layer. These ingredients are sufficient to introduce frustration and to obtain a phase diagram with a spin (or dipolar) glass phase\cite{Vugmeister:1990cr} that may account for the slowness and memory of the polarization process here reported. 

\section{Conclusion}

In this paper we have reported neutron reflectivity measurements performed on BSA molecules adsorbed at silica-water interface with a surface excess near the jamming limit and experiencing an electric field. Depending on the sign of the potential, a second layer is adsorbed or not on top of the first. We argue that the addition of this second layer reveals the slow and remanent electric polarization of the first BSA layer. 
The permanent dipolar structure of BSA suggests an analogy with dipolar glasses that may account for the slowness and memory of the process. We think that this analogy that has not been considered until now opens a new field of investigation, both theoretical and experimental, about adsorbed protein layers dynamics that should give key insights in the understanding of their properties. Actually, in case this analogy were correct, the electric field used in our experiment only reveals the importance of dipole-dipole interactions and frustration that should be present: 1)~even without electrical field; 2)~not only for BSA but more generally for every dipolar proteins and 3)~not only for protein adsorbed at solid-liquid interface but more generally to 2D confined proteins.

%\bibliography{/Users/lairez/Documents/BiblioTex/BiblioDL}

\end{document}